\documentstyle [preprint,revtex]{aps}

\begin{document}
\begin{flushright}
VAND-TH-92-5\\
NSF-ITP-92-77\\
May 1992\\
(Revised September 1992)\\
\end{flushright}
\begin{center}
\LARGE
Wormhole Cosmology and the Horizon Problem
\end{center}
\normalsize
\bigskip
\begin{center}
{\large David Hochberg}\\
{\sl Department of Physics and Astronomy, Vanderbilt University\\
Nashville, TN 37235, USA}\\
and\\
{\large Thomas W. Kephart}\\
{\sl Department of Physics and Astronomy, Vanderbilt University\\
Nashville, TN 37235, USA}
 and
{\sl Institute for Theoretical Physics\\
University of California, Santa Barbara, CA 93106, USA}\\
\end{center}
\vspace{2cm}
\centerline{ABSTRACT}
We construct an
explicit class of dynamic lorentzian wormholes connecting
Friedmann-Robertson-Walker (FRW)
spacetimes. These wormholes can allow two-way transmission
of signals between spatially separated regions of spacetime and could
permit such
regions to come into thermal contact.
The cosmology of a network of early Universe wormholes is discussed.
\vfill\eject

The most convincing evidence for the large-scale smoothness
of the observable Universe is provided by the cosmic microwave
background (CMB) radiation, which was in thermal equilibrium with matter
until the Universe cooled to a temperature of about
4000 $^{\circ}K$, when the radiation
decoupled from matter.  From its
discovery in 1965, numerous measurements have established
the CMB
to be consistent with a blackbody spectrum at a temperature
$T = 2.735 \pm 0.06 {^\circ K}$, and uniform in all
directions (on angular scales from $10''$ to $180^\circ$) to
about a part in $10^4$ [1]. Since this radiation is received
from regions which could not have been in causal contact
at the time of last scattering, the uniformity in temperature
evident in the CMB must be arranged as an initial condition
in the big-bang model [2]. This shortcoming of the
original big-bang model is known as the horizon problem.
There are two ways to solve this problem.  One is to postulate that the
cosmic scale factor underwent an era of exponential expansion
(around 50 e-foldings) such that all the presently observable
Universe is in one inflated causal domain.  This is the inflationary
 Universe scenario which has been exhaustively explored in the
literature [3]. The second possibility is that the
Universe did not necessarily inflate, but has (or had) traversable
bridges (wormholes) connecting otherwise causally disconnected
regions of spacetime. This Universe with ``handles" could alter
and perhaps solve the horizon problem.  Since this approach has
not been studied, we are led to investigate it here, where we
envision a network of traversable wormholes
existing for a very brief time
in the early Universe, arising out of Planck-time
quantum metric fluctuations.

Provided wormholes are important for cosmology, then those
connecting homogeneous and isotropic spacetimes
are undoubtedly the most relevant
ones for study (in contradistinction to astrophysical applications, where
one expects Schwarzschild [4] or Kerr wormholes
to be the most important).
Thus, the first step in any
 investigation of wormhole ``phenomenology," and a
main result of this work, is the construction of
specific cosmological wormhole
solutions to Einstein's equations.  The wormholes we will
consider here result from surgically modified FRW
spacetimes. We adopt this technique for convenience's sake only;
however, we assume the qualitative features of the wormholes are
 independent of
the details of the construction.
To construct them, take two copies of the
FRW solution [5] (adopting units where $c = G =1$)
$$ds^2 = - dt^2 + R^2(t)\left( {{dr^2} \over {1 - \kappa r^2}}
+ r^2(d\theta^2 + sin^2 \theta\, d\phi^2) \right),\eqno(1)$$
and remove from each an identical
four-dimensional region of the form
$\Omega_{1,2} = \{ r_{1,2} \leq a \}$. The resulting
spacetime contains two disjoint boundaries $\partial \Omega_{1,2}
= \{ r_{1,2} = a \}$, which are timelike hypersurfaces.
An orientation preserving identification
$\partial \Omega_1 \equiv \partial \Omega_2$ yields
two FRW spacetimes connected by
a wormhole whose throat is located on their mutual boundary
$\partial \Omega$ [4]. Here, $a = a(\tau)$ is a function describing
the dynamics of the wormhole's throat; the physical radius of the
wormhole is equal to $aR$.
The wormhole is spherically symmetric
and the boundary layer is just the world-volume swept out by
its throat.
This procedure also leads to a wormhole connecting a single
FRW spacetime to itself if one identifies the two background
spacetimes, i.e., we have a FRW space with a handle.
 In this case, the two regions $\Omega_{1,2}$ can
be separated by an arbitrarily large distance in an open Universe.
To insure the modified spacetime is itself a solution of the
gravitational field equations requires a proper matching of the
metric across the boundary layer.
The required junction conditions are most conveniently derived
by introducing gaussian normal coordinates in the neighborhood
of the boundary hypersurface and integrating the Einstein equations across
the boundary [6]. The most general stress-energy tensor
which gives rise to two
identical FRW spaces attached by a wormhole
as described here, is given by
$$T^{\mu}_{\nu} (x) = S^{\mu}_{\nu}\, \delta(\eta) +
{T^{(+)}}^{\mu}_{\nu} \, \Theta (\eta) + {T^{(-)}}^{\mu}_{\nu}
\,\Theta (-\eta),\eqno(2)$$
where ${T^{(+)}}^{\mu}_{\nu} = {T^{(-)}}^{\mu}_{\nu}$ is
one of the standard perfect-fluid source terms leading to (1)
and
$$S^{\mu}_{\nu} \equiv \lim_{\epsilon \rightarrow 0}
\int^{\epsilon}_{-\epsilon} d\eta\, T^{\mu}_{\nu}(x),\eqno(3)$$
is the surface stress-energy. The gaussian coordinate $\eta$
parametrizes the proper distance measured perpendicularly through
$\partial \Omega$. Integrating $G^{\mu}_{\nu} = 8\pi T^{\mu}_{\nu}$
across $\partial \Omega$ and taking the limit as indicated in (3)
yields [6]
$ S^{\eta}_{\eta} = S^{\eta}_{j} = 0,$ and
$$ \left( \Delta {\cal K}^{i}_{j} - \delta^{i}_{j} \,
 Tr(\Delta {\cal K}) \right) = 8\pi S^{i}_{j},\eqno(4)$$
where $\Delta {\cal K}^{i}_{j} \equiv lim_{\epsilon \rightarrow 0}
 ({\cal K^{(+)}}^{i}_{j} - {\cal K^{(-)}}^{i}_{j})$ is the
``jump" in the extrinsic curvature of the hypersurface in going
from the $-\epsilon$ to the $+\epsilon$ ``side".
For the case at hand, reflection symmetry implies $\Delta
{\cal K}^{i}_{j} = 2{\cal K^{(+)}}^{i}_{j}$ and the spherical
symmetry implies ${\cal K}^{i}_{j} = {\rm diag}({\cal K}^{\tau}_{\tau},
{\cal K}^{\theta}_{\theta}, {\cal K}^{\theta}_{\theta})$ and
$S^{i}_{j} = {\rm diag}(-\sigma, P, P)$, where $\sigma, P$ denote
the surface energy and pressure densities.
The extrinsic curvature (in any coordinate system) is defined by
$${\cal K}_{i,j} = n_{\mu} {\bf \nabla}_{(j)} {\rm e}^{\mu}_{(i)},
\eqno(5)$$ where the ${\rm e}_{(i)}$ constitute a set of three
independent tangent vectors defined along the intrinsic coordinates
$\xi^i$ parametrizing the hypersurface, and
$n^{\mu}$ is the outward unit
normal ($n_{\mu}n^{\mu} = +1$, for timelike surfaces);
the background
covariant derivative is taken along the $j^{th}$ coordinate
direction. The throat proper time and the two angles provide a
convenient set of intrinsic coordinates: $\xi^i = (\tau,\theta,\phi)$.
Since the position of the
throat (i.e., its embedding in the background FRW space)
is $X^{\mu} = (t, a(t), \theta, \phi)$, the tangent vectors
are simply ${\rm e}^{\mu}_{(i)} = dX^{\mu}/d\xi^i$, and
the unit normal is $n^{\mu} = ({{a'R} \over {(1 - \kappa a^2)^{1/2}}},
{1 \over R}(1 - \kappa a^2 + (a'R)^2)^{1/2}, 0, 0)$.
A straightforward
calculation of (5) gives
$$ {\cal K}^{\theta}_{\theta} = a'{\dot R} + {1 \over {aR}}
(1 + (a'R)^2)^{1/2},\eqno(6)$$
and
$$ {\cal K}^{\tau}_{\tau} ={{a''R} \over {(1 + (a'R)^2)^{1/2}}}
 + 2 a' {\dot R} ,\eqno(7)$$
where $a'=da/d{\tau}$ and ${\dot R} = dR/dt$ [7].
We have set $\kappa = 0$
since the curvature term in (1) is negligible during
the early stages
of expansion. Note the appearance of the two time parameters
$\tau$ and $t$ in (6) and (7). We can always eliminate one in favor of
the other
(e.g., coordinate time $t$) by using $a' = {\dot a} (dt/d\tau)$
and $$(d\tau/dt) = (1 - ({\dot a} R)^2)^{1/2}.\eqno(8)$$
Physically, this amounts to calibrating the clocks attached to the
throat in terms of the the comoving clocks.
Doing so, and substituting
(6) and (7) into (4) yields the wormhole's equations of motion:
$$-2\pi\sigma = {{{\pm \dot a}{\dot R} + (aR)^{-1}} \over
 {(1 - ({\dot a}R)^2 )^{1/2}}},\eqno(9a)$$
and
$$4\pi P = {  {{\pm \ddot a}R \pm 3{\dot a}{\dot R} + (aR)^{-1}} \over
{ (1 - ({\dot a}R)^2 )^{1/2}}}  \pm
{ {({\dot a}R)^2 [{\ddot a}R + {\dot a}{\dot R}]} \over
{(1 - ({\dot a}R)^2)^{3/2} }}. \eqno(9b)$$
The plus (minus) sign corresponds to expanding (collapsing) wormholes.
We consider
examples of both sign choices below.
The usual redundancy between the Einstein equations and the conservation
of stress-energy $T^{\mu}_{\nu}$, leads to a similar relation between
the wormhole structure equations (9a) and (9b) and the surface stress
tensor (3). The jump in the field equation $G^{\eta}_j = 8\pi T^{\eta}_j$
together with (4) imply [6]
$$\nabla_m S^{m j} = {T^{(-)}}^{\eta j} - {T^{(+)}}^{\eta j} = 0,
\eqno(10)$$
where $\nabla_m$ denotes the covariant derivative intrinsic to the
hypersurface. Thus, the surface stress-energy tensor is conserved.
The derivation of (9a-b) together with (10) constitutes the
most important formal
result of this Letter.

With the wormhole equations of motion at hand, we now
proceed to solve
them.
As a first example, consider the
comoving case, ${\dot a} = a' =  0$. Then
$$\sigma = -{1 \over {2\pi a R}},\eqno(11)$$
and $P = -{1 \over 2}\sigma$.
With a constant throat function $a$, the wormhole's radius simply
grows in direct proportion to the background scale factor $R$.
This behavior is also confirmed by (8), which implies
$t = \tau$. These two time scales coincide if and only if the
wormhole throat is coupled to the expansion.
The surface energy density is negative,
as indeed it must be: the violation of the weak energy condition
(WEC) at the wormhole throat is expected on general grounds.
In simple terms, the spacetime region surrounding the throat
acts as a diverging lens which defocusses light and particle
geodesics as they traverse the wormhole. This path divergence
reflects a gravitational repulsion at the throat and signals
the presence there of a localized
negative energy density. Although no
classical stress energy tensor is known to violate the WEC, quantum
fields in curved backgrounds tend to develop localized negative
energy densities [9,10], and suggests these wormholes are
driven by quantum field fluctuations, their back-reaction
on the metric [11]
and by the quantum fluctuations in the metric, as
envisioned, for example, in the spacetime foam [12].
According to (11), the comoving
wormholes can exist for any value
of the fixed throat function $a$.
The magnitude of the surface energy density decreases for
increasing scale factors.
While these comoving wormholes do solve the structure
equations, it may be more likely for FRW wormholes to evolve
in time in such a way that they decouple from the expansion.
Thus we are led to consider decoupled
dynamic wormholes, where the full nonlinearity of the structure
equations (9a-b) comes into play.

The strategy for solving the general time
dependent case
involves selecting a relevant scale factor $R$,
a surface energy density $\sigma$ and an equation of state
$P = P(\sigma)$ subject to (10).
First let us assume a power law expansion
$R(t) = R_i \left({t \over {t_i}} \right)^{p}$. Then
 the choice for $a(t)$ that most simplifies (9a) and (9b) is
$a(t) = a_i \left({t \over {t_i}} \right)^{1-p}$
which renders all the denominators constant.
Here $R_i$ and $a_i$ are the scale factor and throat
function at some initial time $t_i$.  Now both $P$ and $\sigma$
go like $t^{-1}$ and the equation of state is in general nonlinear.
The relevant power law for early Universe applications is $p = 1/2$,
i.e., a radiation dominated expansion.
For this case the  surface energy density
is  $\sigma = \epsilon_0/a^2$, where
$\epsilon_0$ is a constant with units
of mass given below.
Then the \underline{exact}
solution of (9a) and (9b) has
$$a(t) = a_i \left( {t \over {t_i}} \right)^{1/2},\eqno(12)$$
and an equation of state as given below.
Note that
the wormhole throat function
is related to the background expansion:
${\dot a}/a = {\dot R}/R$. Moreover, the wormhole motion is decoupled
from the expansion since
$ {1 \over {aR}} {{d(aR)} \over {dt}}  > {{\dot R} \over R}$.
The connection between proper and cosmic time implied by (8) is
$t = const. \times \tau$, where the constant of proportionality
is greater than one, and depends on $a_i, R_i$ and $t_i$.

 For each $t_i$, there is an upper
bound on the size spectrum of initial wormhole throat radii which
follows if we assume no point on the throat can move faster
than the speed of light [13]: $|\beta| < 1$ where
$\beta \equiv  2{\dot a} R/c $ is a constant for the class of
solutions treated here (with $p=1/2$).
For radiation dominated expansion, this condition implies
$$a_i R(t_i) \leq (aR)_{max} \equiv c\,t_i ,\eqno(13)$$
where $R(t_i)$ is the scale factor for radiation dominance
evaluated at some initial time $t_i$ [14].
The surface energy density is given by
(restoring $c$ and $G$)
$$\epsilon_0 = - {{c\, {a_i}^2} \over {4\pi G\, t_i}}
\left( { {\beta + \beta^{-1}} \over {(1 - \beta^2)^{1/2}}} \right)
\eqno(14)$$
and is negative, as for the comoving case.

The equation of state for the $p = 1/2$
case is
given by
 $$P = -{{\sigma } \over {2}}
{\left( {1 \pm 2\,{ \beta}^2} \over {1 \pm \beta^2} \right)}, \eqno(15)$$
with sign choices as in (9a) and (9b).
Note specifically that in the $\beta \rightarrow 0$ limit $P=- \sigma /2$
which corresponds to comoving wormholes in agreement
with (11) and the subsequent discussion.  Other interesting limits are
$\beta = 1$ for expanding wormholes where $P = -3{\sigma}/4$;
while
for collapsing wormholes, $\beta = 1/{\sqrt{2}}$
yields $P = 0$,
corresponding to
pressureless negative energy ``dust"; and $\beta \rightarrow 1$
where $P$ diverges.

Thus far, we have dealt with aspects of a single
comoving or decoupled
wormhole in a FRW background. It is clear, however, that the above
construction can be applied an indefinite number of times and will
lead to an explicit solution of Einstein's equation with an
arbitrary number of wormholes.
In fact, this construction is an application
of the connected sum operation, familiar from algebraic topology.
In general, we let ${\cal M}_1$ and ${\cal M}_2$ be two
spacetimes and remove from each the interiors of the four-dimensional
discs $D_1 \subset {\cal M}_1$ and $D_2 \subset {\cal M}_2$, and
then identify the boundary sets up to homeomorphism
$h : \partial D_1 \rightarrow \partial D_2$. The resulting manifold
defines the connected sum of ${\cal M}_1$ and ${\cal M}_2$:
${\cal M}_1 \# {\cal M}_2$ [15]. The operation $\#$ is both
commutative and associative. This latter property implies that
{\it multi-wormhole} solutions of Einstein's equation can be
constructed unambiguously.

Let us end with a discussion (and some speculations on) the applications
of our results.
To be of interest, wormholes need only
stay open long enough for the radiation to traverse the throat.
For the class of solutions treated here, a time scale is set
by the frequency of the radiation traversing the throat,
 $ \Delta t \sim 1/{\nu}$, and a length
scale by $\lambda \sim a_i R_i $.
This follows since our wormholes have
zero-length throats, that is, a particle going down the wormhole mouth
in one region comes out the other end instantaneously
(see Fig.1).
 Of course, we do not claim all wormholes will be of this
form, but this still serves as a useful estimate
of the minimum wormhole lifetime needed to initiate thermal contact
at a fixed frequency.
With these reservations [17] we now procede with a purely illustrative
example. We assume a number density $n(t_{Pl})$ of Planck-sized
wormholes (of radius $l_{Pl}$) at the Planck time [18], and we let
$n(t_{Pl}) = \gamma\, n_{Pl}$ ($\gamma = {\rm const.}$) [19]
where
$n_{Pl} = (l_{Pl})^{-3}$ is the Planck number density. At some
later time $t$, the wormhole number density is
$n(t) = \gamma\, n_{Pl} (R_{Pl}/R(t))^{3}$. The volume of space filled
by the interior of wormhole mouths (volume filling factor) is
$$\phi(t) = { {n(t)} \over {n_{Pl}} }
({ {l(t)} \over {l_{Pl}} })^{3}.\eqno(16)$$
For comoving wormholes $\phi(t) = \gamma$, as expected, but let us concentrate
on dynamic $p = 1/2$ wormholes where $l = aR$, for which one finds
$\phi(t) = \gamma (R(t)/R_{Pl})^3$. Now for example, the average
particle in the cosmic soup will have traversed at least one wormhole
by time $t$ if $\gamma > (R(t)/R_{Pl})^3$. For example, by the
GUT time,
$t_{GUT} \sim 10^{-34} {\rm sec}$, a $\gamma \geq 10^{-21}$ allows
thermalization.
Thermalization at even the high $\nu$ end of the spectrum can
(small wormholes only)
eventually lead to thermal equilibrium at a later time
 for causally disconnected regions. This in turn could provide a
wormhole solution to the horizon problem.  It is rather natural to
have an inflationary era in the early Universe.
There may also be a cosmic wormhole era in the early Universe.
However, to compete with inflation, a
wormhole cosmology needs to address a list of other cosmological
problems including flatness, monopoles (this may be solved directly
in certain particle physics models) and density fluctuations.  All
these need further study.

TWK thanks the Institute of Theoretical Physics at Santa Barbara
for hospitality while this work was in progress.  This work was
supported by the Department of Energy under Grant No.
DE-FG05-85ER40226, and by the National Science Foundation under
Grant No. PHY89-04035.

\vspace{1.5cm}
\begin{itemize}
\item{1.} The status of early CMB measurements is reported in
{\it After the First Three Minutes}, edited by S.S. Holt,
C.L. Bennett and V. Trimble (American Institute of Physics, 1991).
A dipole anisotropy has been previously seen in the COBE data, and
recently evidence for higher multipole moments at the level of
a few parts in $10^{5}$ has been reported in: ``Interpretation of the CMB
Anisotropy Detected by the COBE DMR,"   E. L. Wright et al., COBE preprint
Apr. (1992), and ``Structure in the COBE DMR First Year Maps",
G. F. Smoot et al., COBE preprint Apr. (1992).
\item{2.} For a review of the standard big bang model, see, e.g.,
E.W. Kolb and M.S. Turner, {\it The Early Universe},
(Addison-Wesley, New York, 1990).
\item{3.} A.H. Guth, Phys. Rev. {\bf D}23, 347 (1981);
A.D. Linde, Phys. Lett. {\bf B}108, 389 (1982);
A. Albrecht and P.J. Steinhardt, Phys. Rev. Lett. {\bf 48}, 1220 (1982).
For a review of inflationary cosmology, see e.g., Ref.  [2],
Chap. 8; K. A. Olive, Phys. Rep. {\bf 190}, 307 (1990),
and references therein.
\item{4.} The construction we employ here has been
used to study wormholes in
Minkowski, Schwarzschild and Nordstr\"om-Reissner backgrounds by
M. Visser, Phys. Lett. {\bf B}242, 24 (1990); Nucl. Phys. {\bf B}
328, 203 (1989).
\item{5.} The scale factor $R$ is dimensionless; the curvature
constant $\kappa = -1, 0$ or 1, corresponding to negative, flat or
positively curved spatial sections, respectively.
\item{6.} W. Israel, N. Cimento 44{\bf B}, 1 (1966); 48{\bf B}, 463
(1967); C.W. Misner, K.S. Thorne and J.A. Wheeler,
{\it Gravitation}, (Freeman, San Francisco, 1973), Chap 21.13.
\item{7.} The metric on the throat world-volume, $^{(3)}g_{ij} =
{\rm diag}(g_{\tau \tau}, g_{\theta \theta}, g_{\phi \phi})$,
has components $g_{\tau \tau} = -1$, $g_{\theta \theta} =
a^2(\tau) R^2(t)$ and $g_{\phi \phi} = sin^2 \theta\, g_{\theta \theta}$.
\item{8.} M.S. Morris and K.S. Thorne, Am. J. Phys. {\bf 56}, 395
(1988); M. Morris, K.S. Thorne and U. Yurtsever, Phys. Rev. Lett.
{\bf 61}, 1446 (1988).
\item{9.} N.D. Birrell and P.C.W. Davies,
{\it Quantum Fields in Curved Space}, (Cambridge U.P., Cambridge, 1982).
\item{10.} D. Hochberg and T.W. Kephart, Phys. Lett. {\bf B}268, 377
(1991).
\item{11.} D. Hochberg and T.W. Kephart, VAND-TH-92-3, March 1992.
\item{12.} A quantum mechanical analysis of Minkowski signature
wormholes in static background spacetimes
has been carried out by M. Visser, Phys. Rev. {\bf D}43, 402
(1991).
\item{13.} Though we adhere to $|\beta| \leq 1$, this may not
be required since the matter corresponding to
the negative energy density on
the wormhole throat may behave tachyonically.
\item{14.} The scale factor for a radiation dominated Universe is
$R(t) = (32\pi G\, \rho_{rad}/3)^{1/4}\,t^{1/2}$,
where $\rho_{rad} = 4.5 \times 10^{-34} {\rm gm/cm^3}$ is the present
radiation energy density.
\item{15.} See, e.g., W.S. Massey, {\it Algebraic Topology},
(Springer, Berlin, 1977). Our wormholes result from the trivial
homeomorphism, $h \equiv {\rm identity}$.
\item{16.} D. Hochberg, Phys. Lett. {\bf B}251, 349 (1990).
\item{17.} It has been suggested that wormholes may form time machines
(see second paper in Ref. [8]); if this were the case, the time travel
aspects of wormholes could drive the Universe away from thermalization.
However, based on quantum mechanical arguments, Hawking has
argued (Phys. Rev. {\bf D}46, 603 (1992)) and
Visser has given a partial proof (WASH-U-HEP-92-70, Feb. 1992)
that wormholes cannot be used for time-travel.
Hence we assume
travel through these wormholes always leads the Universe towards
equilibrium.
\item{18.} The weak energy condition (WEC) is a classical conjecture
that requires
the energy density at each point in
spacetime to be nonnegative.
For an energy fluctuation $\Delta E \sim |\sigma a^2 R^2|$, we
find that Planck-sized wormholes of age $\Delta t \sim t_{Pl}$
only violate the WEC at the quantum fluctuation level
$\Delta E \Delta t \sim \hbar$, and so these wormholes cannot be
ruled out by the WEC arguments. A longer-lived wormhole (say, a growing
$p=1/2$ wormhole) will soon have $\Delta E \Delta t \geq \hbar$.
Hence if the WEC is indeed applicable here, it will be violated, apparently
implying the wormhole's eventual recollapse. We as yet do not know
how to make these observations more quantitative.
\item{19.} Constant $\gamma$ implies the total number of wormholes is
constant. This need not be the case in general, and we expect
$\gamma = \gamma(t)$.
\end{itemize}

\vfill\eject
%\vspace{1.5cm}

\begin{center}
{\bf FIGURE CAPTION}
\end{center}
\noindent
Figure 1. A spacetime diagram representation
of a Friedmann cosmology with flat ($E^{3}$) spatial sections, with
the past lightcone for an observer at $A$.  Radiation
received at $A$
from opposite directions was emitted from $B$ and $C$.  A wormhole
connects events in the past lightcones of the otherwise spatially
seperated points $B$ and $C$.  A photon is shown going down the wormhole
in $B$'s past and exiting the wormhole in $C$'s past.

\end{document}